\pdfoutput=1
\documentclass[amsmath,amssymb,aps,twocolumn,superscriptaddress]{revtex4-2}
\usepackage[utf8]{inputenc}
\usepackage[T1]{fontenc}
\usepackage[english]{babel}

\usepackage{amsmath}
\usepackage{amssymb}
\usepackage{mathtools}
\usepackage{bm}
\usepackage{braket}
\usepackage{physics}
\usepackage{MnSymbol}
\usepackage{wasysym}
\usepackage{cancel}

\usepackage{graphicx}
\usepackage{epsf}
\usepackage{xcolor}
\usepackage[version=4]{mhchem}
\usepackage{siunitx}

\usepackage{caption}
\usepackage{subcaption}
\usepackage[bookmarks=false]{hyperref}
\usepackage{natbib}

\usepackage{dcolumn}
\usepackage{tabularx}
\usepackage{array}
\usepackage{booktabs}

\usepackage{float}
\usepackage{placeins}
\usepackage[shortlabels]{enumitem}
\usepackage{lipsum}

\makeatletter
\def\maketitle{
\@author@finish
\title@column\titleblock@produce
\suppressfloats[t]}
\makeatother
\newcommand{\beginsupplement}{
    \setcounter{table}{0}
    \renewcommand{\thetable}{S\arabic{table}}
    \setcounter{figure}{0}
    \renewcommand{\thefigure}{S\arabic{figure}}
    \setcounter{equation}{0}
    \setcounter{section}{0}
    \renewcommand{\theequation}{S\arabic{equation}}
}

\begin{document}
\title{Controlled Switching of Bose-Einstein Condensation in a Mixture of Two Species of Polaritons}

\author{Hassan Alnatah}
\affiliation{Joint Quantum Institute, University of Maryland, 4254 Stadium Dr., College Park, Maryland 20742, USA}

\author{Shuang Liang}
\thanks{H.A. and S.L. contributed equally to this work.}
\thanks{address correspondence to: SHL450@pitt.edu}
\affiliation{Department of Physics, University of Pittsburgh, 3941 O’Hara Street, Pittsburgh, Pennsylvania 15218, USA}

\author{Qiaochu Wan}
\affiliation{Department of Physics, University of Pittsburgh, 3941 O’Hara Street, Pittsburgh, Pennsylvania 15218, USA}

\author{Jonathan Beaumariage}
\affiliation{Department of Physics, University of Pittsburgh, 3941 O’Hara Street, Pittsburgh, Pennsylvania 15218, USA}

\author{Kirk Baldwin}
\affiliation{Department of Electrical Engineering, Princeton University, Princeton, New Jersey 08544, USA}

\author{Adbhut Gupta}
\affiliation{Department of Electrical Engineering, Princeton University, Princeton, New Jersey 08544, USA}

\author{Loren N. Pfeiffer}
\affiliation{Department of Electrical Engineering, Princeton University, Princeton, New Jersey 08544, USA}

\author{David W. Snoke}
\affiliation{Department of Physics, University of Pittsburgh, 3941 O’Hara Street, Pittsburgh, Pennsylvania 15218, USA}

\date{\today}

\begin{abstract}
We report temperature-dependent switching between lower and upper polariton condensation in a GaAs/AlGaAs microcavity when both of these species have comparable populations in a mixture. Using angle-resolved photoluminescence, we observe that at low temperatures, condensation occurs in the lower polariton branch, while at elevated temperatures, the upper polariton branch can become favored. At an intermediate temperature, we observe instability in the condensate formation, characterized by metastable correlations of the fluctuations in intensity and linewidth of the lower and upper polariton branches. 

\end{abstract}

\maketitle

\section{INTRODUCTION}
Exciton-polaritons are quasiparticles formed by the strong coupling of semiconductor quantum well excitons and cavity photons \cite{weisbuch1992observation}. Polaritons can be viewed as photons dressed with an effective mass and repulsive interactions, making them ideal candidates for studying Bose-Einstein condensation (BEC) \cite{deng2002condensation,kasprzak2006bose,balili2007bose,abbarchi2013macroscopic,sanvitto2010persistent,lagoudakis2009observation,amo2009superfluidity,nardin2011hydrodynamic} and for nonlinear optical devices \cite{sanvitto2016road,ballarini2013all,amo2010exciton}.
\par
Over the past two decades, polariton condensation has been primarily observed in the lower polariton (LP) branch, where favorable relaxation dynamics allow efficient scattering to the ground state \cite{tassone1999exciton,porras2002polariton}. In contrast, the upper polariton (UP) branch, which is at higher energy, has rarely been associated with condensation, primarily due to low thermal population at low temperatures and the efficient conversion of UP to LP by emitting optical phonons.
Recent experimental advances, however, have enabled the observation of condensation in the UP branch under specific conditions ~\cite{chen2023bose}. This result challenges the conventional assumption that polariton condensation is strictly confined to the LP state, and opens the door to exploring the complex landscape of condensate mode competition. 
\par
Recent theoretical work has predicted that under non-equilibrium conditions, condensation can occur in the UP branch due to a non-Hermitian phase transition between polariton BEC and photon lasing~\cite{hanai2019non}. In particular, Hanai et al.~showed that bistability and switching between LP and UP condensates can occur near an exceptional point. Our results provide direct experimental evidence for this behavior, revealing temperature-induced switching and temporal fluctuations between LP and UP condensation near threshold. In many ways, this work is also analogous to BEC in multi-species systems in cold atoms gases (e.g., Refs.~\cite{frapolli2017stepwise,stamper2013spinor,myatt1997production}), but with free interconversion between the species.    
\par
Mode competition and switching have been studied extensively in driven-dissipative polariton condensates, including optical bistability \cite{baas2004optical}, polariton multistability \cite{gippius2007polarization,paraiso2011erratum}, fluctuation-driven switching near bistable points \cite{rodriguez2017probing}, critical fluctuations near the condensation threshold \cite{alnatah2024critical}, and temporal fluctuations of $g^{(2)}$ near threshold characterized by stochastic hopping between condensed and non-condensed states \cite{rozas2026temporal}. Here, we report a different kind of switching: not between two states of the same branch, but between two condensate states on two distinct polariton branches.
\par
In this work, we use the temperature of the cryostat bath to tune switching between condensation in the LP and UP branches in a GaAs/AlGaAs microcavity system. By tuning the lattice temperature, we observe that LP condensation is favored at low temperatures and switches to UP condensation at high temperatures. At a specific intermediate temperature around 80 K, the condensate displays stochastic switching behavior, alternating between the two branches even under identical experimental conditions. This behavior suggests that the system enters a regime of competing condensate states near the condensation threshold, where small fluctuations in carrier dynamics or relaxation pathways can determine the dominant condensate mode.
\par
In addition to its fundamental interest, controllable switching between lower- and upper-polariton condensation has direct implications for polariton-based photonic devices. The ability to select the condensate branch using a temperature control provides a new mechanism for realizing bistable coherent light sources in a medium with strong nonlinearity. 

\newpage
\section{EXPERIMENTAL METHODS}

In the experiments reported here, we used a GaAs/AlGaAs microcavity structure very similar to those of previous experiments \cite{alnatah2024bose,alnatah2025strong,liang2025mirror}.  The microcavity sample consisted of a total of 12 GaAs quantum wells with AlAs barriers embedded within a distributed Bragg reflector (DBR). The DBRs are made of alternating layers of AlAs and \ce{Al_{0.2}Ga_{0.8}As}. The top DBR is composed of 23 pairs and the bottom DBR is composed of 40 pairs. The quantum wells are in groups of four, with each group placed at one of the antinodes of cavity. A schematic illustration of the sample structure is provided in the Supplementary Material. A key feature of the experiments described here is that they were performed using a sample with a lower Q factor than the highest-quality samples available. For the very high-$Q$ samples used in previous works (e.g., \cite{steger2015slow,nelsen2013dissipationless,alnatah2024coherence}), the scattering from UP to LP is faster than the decay rate, making the UP hard to detect. As the cavity lifetime is decreased, it becomes comparable to the time for conversion from UP to LP, so that we can easily see the dynamics of the two populations competing for condensation. We therefore used a variation with $Q$ nominally equal to $\sim 4\times 10^4$.
\par
The sample was placed in a variable-temperature microscope cryostat, and
the polaritons were generated by non-resonantly exciting the sample with a wavelength-tunable continuous wave laser, set near a reflectivity minimum (719.2 nm). To minimize heating of the sample, the pump was modulated using an optical chopper operating at a 1$\%$ duty cycle, producing pulses of roughly 25 $\mathrm{\mu s}$--much longer than the intrinsic timescales of the polariton dynamics. The non-resonant pump excitation created electron-hole pairs, which relaxed in energy and formed polaritons. The photoluminescence (PL) was collected using a microscope objective with a numerical aperture of 0.75 and was imaged onto the entrance slit of a spectrometer. The image was then sent through the spectrometer to a CCD camera for time-integrated imaging. 
\par
To measure the energy dispersion of the polaritons, we used angle-resolved photoluminscence (PL) to obtain the intensity image $I(\theta,E)$, where $\theta$ is the angle of emission, which has a one-to-one mapping to the in-plane momentum of the polaritons. The angle of photon emission maps directly to the in-plane $k$-vector of the particles inside the structure \cite{carusotto2013quantum,deng2010exciton,houdre1994measurement}.

\section{EXPERIMENTAL RESULTS}
\begin{figure}
\includegraphics[width=1\columnwidth]{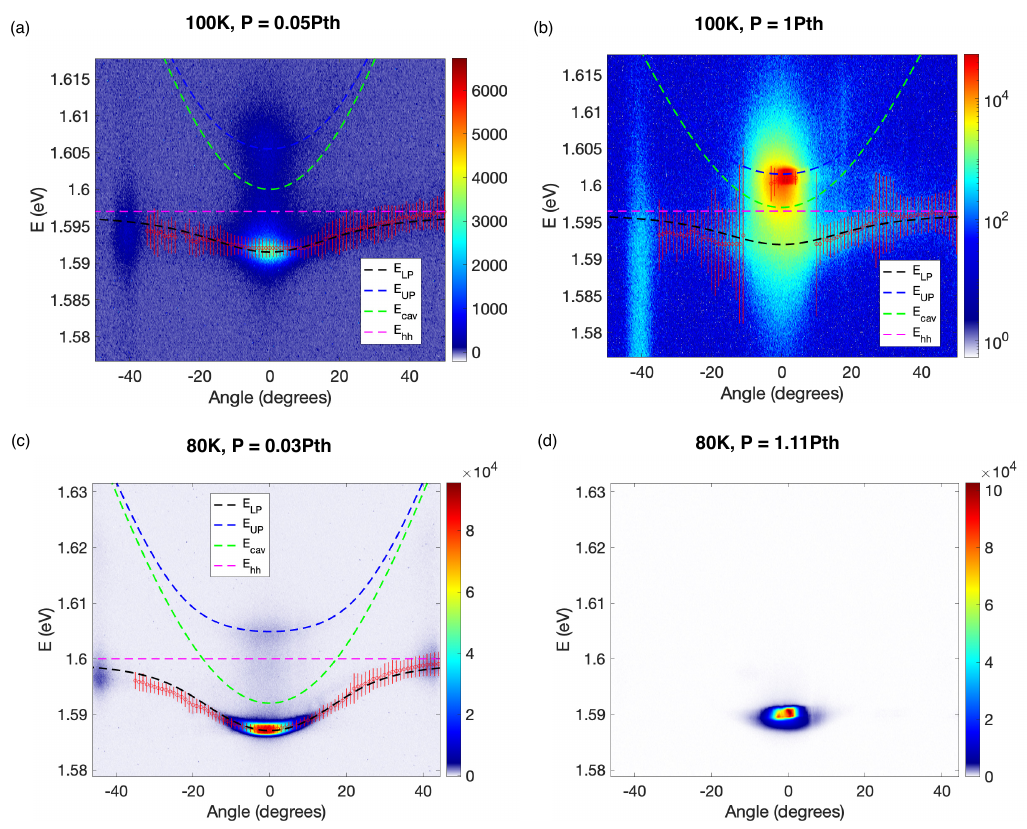}
\centering
\caption{\textbf{Energy-resolved PL at 100 K and 80 K.} {\bf (a)} angle-resolved PL measurements of the polariton at very low pumping power with a LP exciton fraction of 0.61 and {\bf (b)} high pumping power at 100 K. (c,d) angle-resolved PL measurements at 80 K and LP exciton fraction 0.27 under low (c) and high (d) pumping conditions.}
\label{fig:EvsK_condensation}
\end{figure}

Figure \ref{fig:EvsK_condensation} shows examples of dispersion of the polaritons measured at two different temperatures, collected through a pinhole placed at the center of the pump spot to select a spatially homogeneous region, clearly demonstrating strong coupling with two distinct polariton branches. These branches, lower-polariton (LP) and upper-polariton (UP), arise due to the interaction between the heavy-hole exciton with the cavity photon mode. (At high temperature, the light-hole exciton comes into play, as discussed in Ref.~\cite{alnatah2025strong}, but here can be ignored.) The real-space images, the pinhole diameters, and further details are given in the Supplemental Material.
\par
As the pump power is increased across the condensation threshold, polariton condensation was observed. Figure \ref{fig:EvsK_condensation} shows angle-resolved PL at low and high pumping powers for each temperature. Consistent with typical condensation behavior, the PL narrows both in $k$-space and in energy width as the density increases; the latter indicates temporal coherence of the emission \cite{richard2005experimental,kasprzak2008second,roumpos2012power,brune2025quantum}. Further quantitative experimental results for the PL lines is presented in the Supplementary Material (Figs. S2 and S3).

\begin{figure}
\includegraphics[width=0.7\columnwidth]{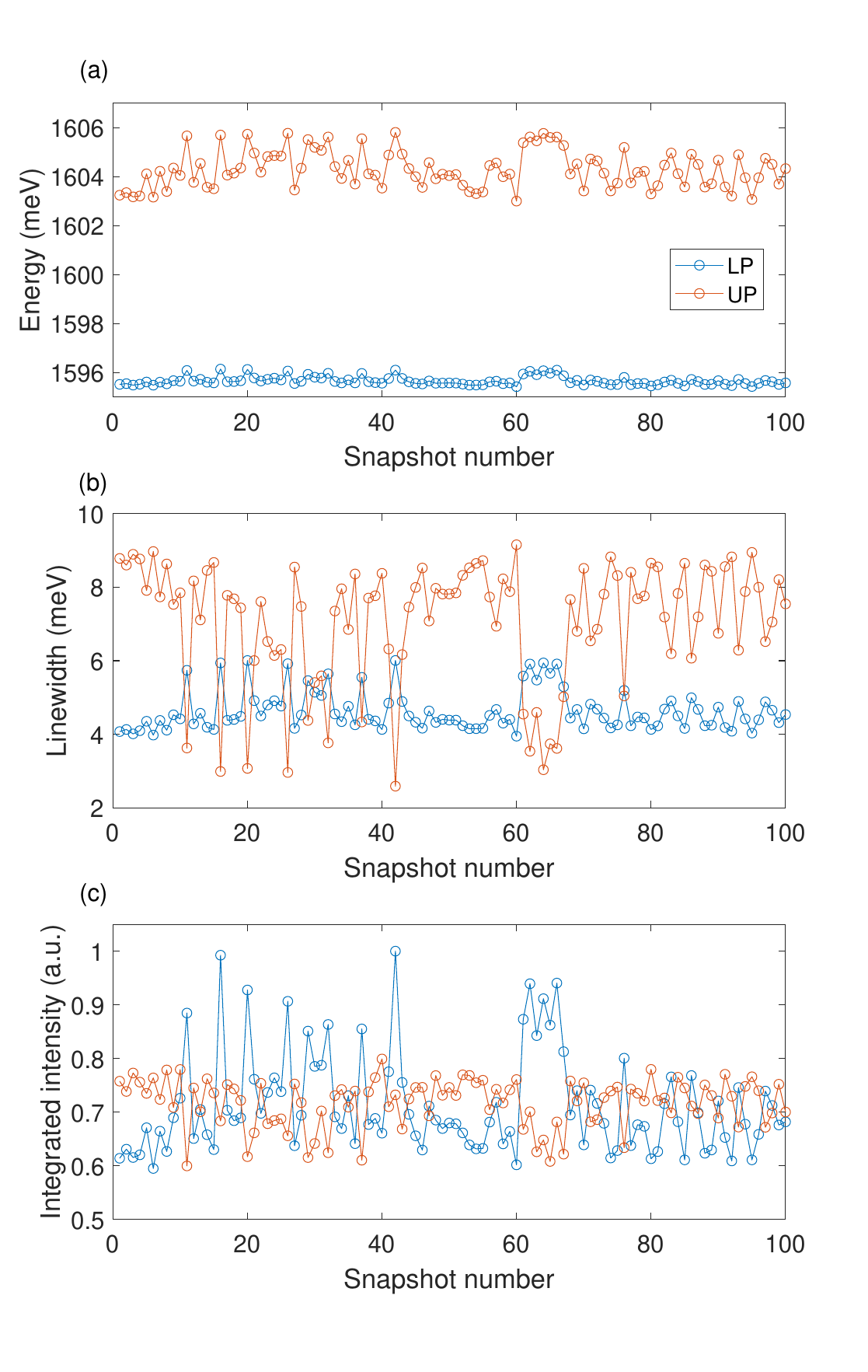}
\centering
\caption{\textbf{Anticorrelated coherence at a fixed power and temperature (80 K)}, for LP exciton fraction of 0.67, near the condensation threshold. {\bf (a)} The energy, (b) the linewidth and (c) the maximum intensity as a function on snapshot number.}
\label{fig:switching_dynamics}
\end{figure}
\par
We find that branch which condenses first depends on both the temperature and the exciton fraction. For example, at a lattice temperature of 100 K, the polaritons condense in the upper polariton branch when the exciton fraction of the LP is {0.61}, as shown in Fig. \ref{fig:EvsK_condensation}(b). In contrast, at 80 K and a LP exciton fraction of {0.27}, condensation occurs in the lower branch, as seen in Fig. \ref{fig:EvsK_condensation}(d). This temperature-dependent switching between upper and lower branch condensation highlights the sensitivity of the system to thermal and relaxation dynamics, as well as the exciton fraction which sets the masses of both branches. For these conditions, the branch favored for condensation is determined by the combined effects of temperature and detuning, the latter setting the exciton fractions and therefore the effective masses of the two polariton branches. 

\par
Notably, for some values of temperture and exciton fraction (in this case $80 \pm 2$~K and exciton fraction of approximately 0.67), the polariton system exhibits an instability in mode selection precisely at the condensation threshold. That is, the polaritons fluctuate between condensation in the UP and LP branches even while the experimental conditions remain constant, as if the polaritons cannot “decide” which state to macroscopically occupy.
\par
To study this behavior, we took a series of snapshots of the PL, where each snapshot was time integrated for 100 ms and represents an average over 40 laser pulses. Despite identical experimental conditions across snapshots, the system displays fluctuations in both linewidth and intensity, alternating between the LP and UP branches as shown in Fig. \ref{fig:switching_dynamics}. Interestingly, the system shows an an anticorrelated switching behavior, where the linewidths and integrated intensity of the LP and UP branches vary in opposite directions.

\begin{figure}
\includegraphics[width=0.7\columnwidth]{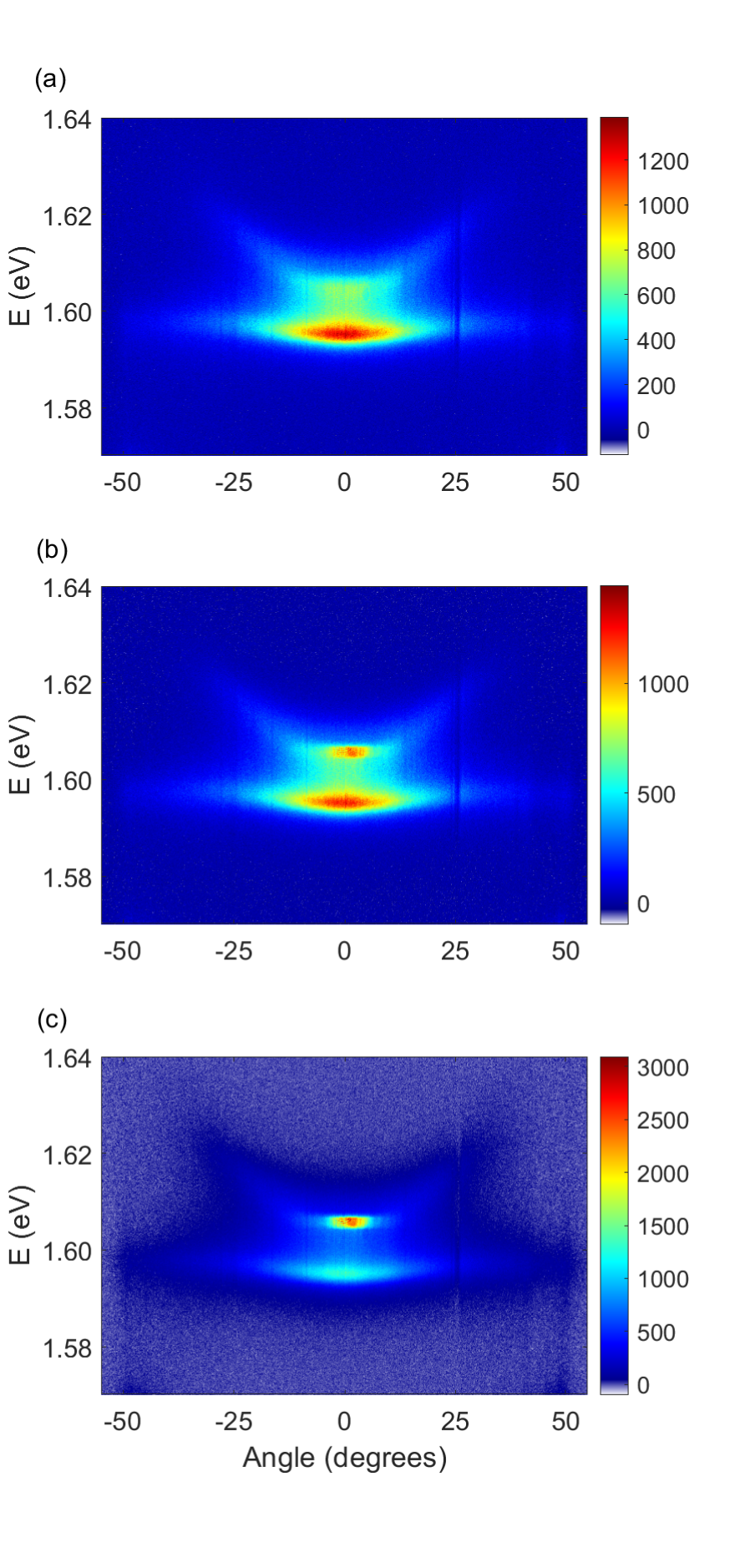}
\centering
\caption{\textbf{Selected snapshots at fixed temperature of 80 K}, for LP exciton fraction of 0.67 and pump power corresponding to Fig.~\ref{fig:switching_dynamics}, for identical experimental conditions. {\bf (a)} The LP is brighter, {\bf (b)} the LP and UP have the same brightness, {\bf (c)} the UP is brighter.}
\label{fig:switching_images}
\end{figure}

\par
Figure \ref{fig:switching_images} shows selected angle-resolved photoluminescence images taken under identical experimental conditions, illustrating the underlying fluctuations that give rise to the statistical data shown in Fig. \ref{fig:switching_dynamics}. Each panel represents a single 100 ms time-integrated snapshot and highlights a different realization of the condensate behavior. In some instances, the emission is dominated by the lower polariton (LP) branch (Fig.~\ref{fig:switching_images} (a)), indicating that condensation has occurred predominantly in the LP state. In other cases, the upper polariton (UP) branch is significantly brighter (Fig. \ref{fig:switching_images} (c)), suggesting that the condensate has formed in the higher-energy UP state, and in some cases, both branches exhibit comparable brightness, indicating either rapid switching between the two which is averaged in our time-integrated measurements, or coexistence between LP and UP condensation (Fig. \ref{fig:switching_images}(b)).

We note that the condensate does not have the same degree of coherence in the LP and UP branches when this instability occurs. As seen in Fig.~\ref{fig:switching_dynamics}, the UP condensate, when it exists, has quite narrow spectral line, while the LP condensate has a typical full width at half maximum of 4 meV. This is narrower than its width when the UP is condensed, indicating that the LP in this case is a nonequilibrium condensate-like state with an increased degree of coherence, but is not a fully coherent condensate. 
Also, we note that while the UP condensate is narrower, its energy position is more unstable.

\begin{figure}
\includegraphics[width=0.7\columnwidth]{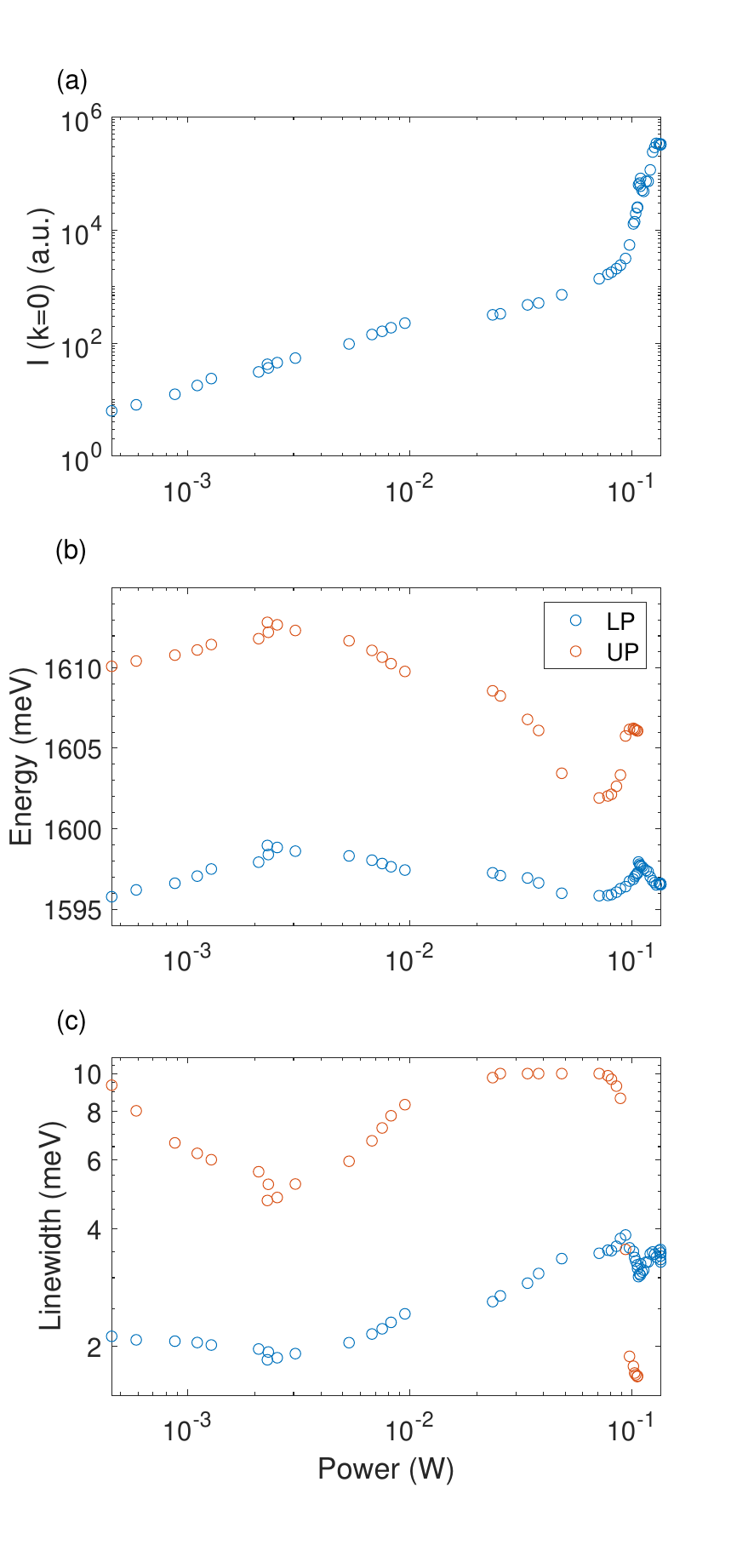}
\centering
\caption{\textbf{Characteristics of the PL lines at 80 K}, averaged over many shots as in the data of Figure~\ref{fig:switching_dynamics}. \textbf{(a)} The intensity at $k=0$ of the polaritons as a function of pump power. \textbf{(b)} The energies of the polariton lines at $k=0$ as a function of the pump power. \textbf{(c)}  Full width at half maximum at $k=0$.}
\label{fig:condensation}
\end{figure}

\par
To further quantify the condensation behavior, we extracted the average linewidth and intensity by integrating the PL signal over many laser pulses (typically more than 480) at a temperature of 80 K, corresponding to the same data as in Figs.~\ref{fig:switching_dynamics} and \ref{fig:switching_images}. In these plots, the data from multiple shots is averaged. Near the condensation threshold, a nonlinear increase in intensity is observed. This sharp increase in intensity is accompanied by a decrease in the line width of the UP by approximately a factor of 10.

\begin{figure}
\includegraphics[width=1\columnwidth]{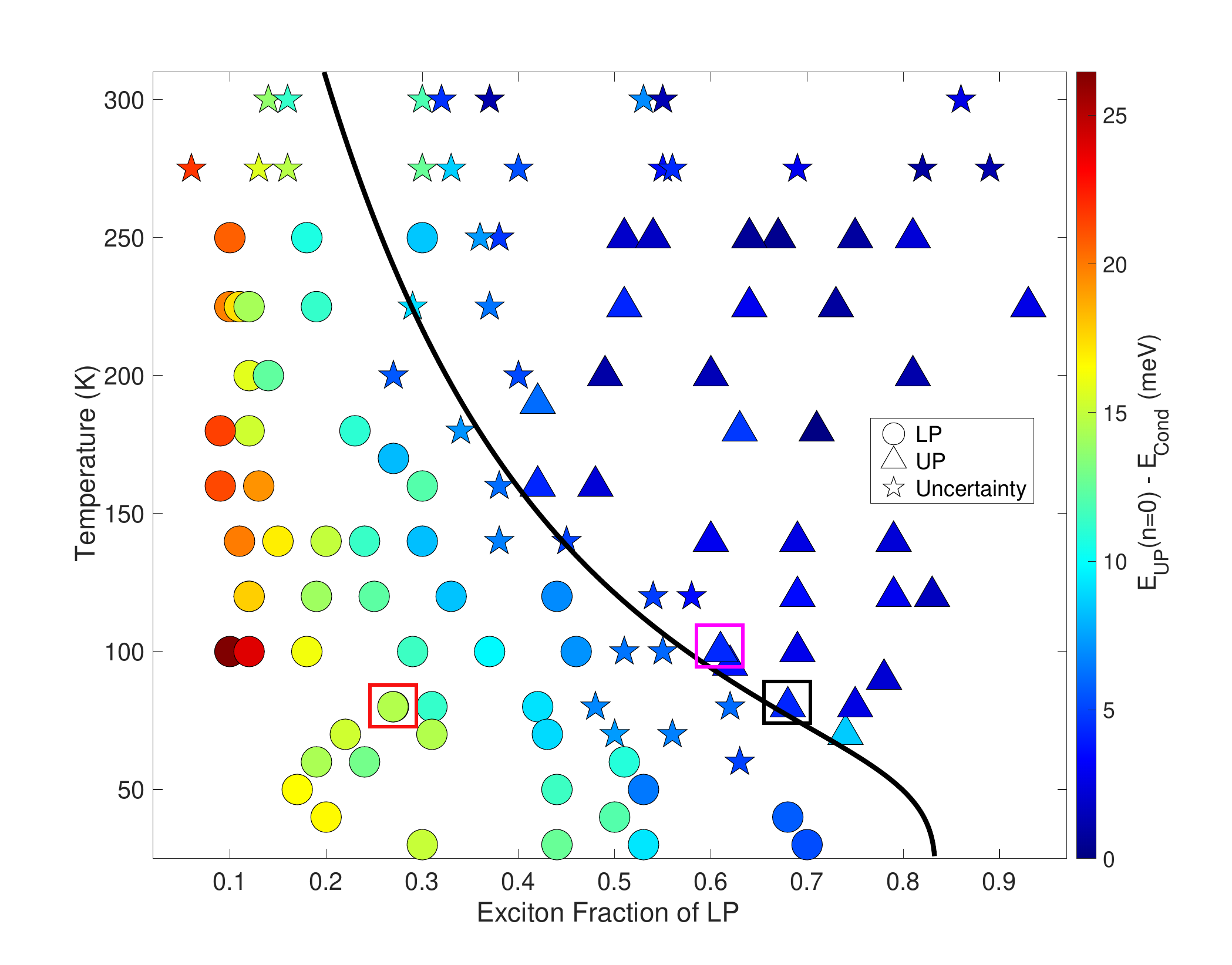}
\centering
\caption{\textbf{Phase diagram} showing the LP exciton fraction versus temperature for LP and UP condensates. The exciton fraction for various locations on the microcavity sample has been deduced using fits to the energies $E_{LP}(\theta)$ and $E_{UP}(\theta)$ from data like that shown in Figure~\ref{fig:EvsK_condensation} at low pump density, using a model like that of Eq.~(\ref{matrix}) but with three levels for the photon, heavy-hole exciton and light-hole exciton in GaAs quantum wells.
Triangles represent conditions where condensation occurs in the upper polariton branch, while circles represent condensation in the lower polariton branch. The color bar gives the emission energy of the condensate relative to the zero-density upper polariton branch. The squares identify conditions shown elsewhere in this paper: the magenta square corresponds to Fig.~\ref{fig:EvsK_condensation}(a,b), the red square to Fig.~\ref{fig:EvsK_condensation}(c,d), and the black square to the metastable jumping between the two branches seen in Figs.~\ref{fig:switching_dynamics} and \ref{fig:switching_images}. The solid black line is given by Eq.~\eqref{eq:boundary} for the parameters given in the text.}
\label{fig:phase_diagram}
\end{figure}

Finally, in Figure~\ref{fig:phase_diagram}, we plot a nonequilibrium-state phase diagram. All of the symbols represent conditions in which we can clearly identify the upper and lower polariton energies at low density. The shape of the symbol indicates whether the condensate, when it appears, occurs in the lower or upper polariton state. Stars indicate conditions in which we can see only one line when the system is a condensate. The color of the symbols gives the energy of the condensate relative to the upper branch energy at low density. As seen in this figure, at high temperatures, the condensate forms in whichever branch is more photonic, and therefore has the lighter effective mass of the two branches. When the lower polariton is photon-like (low exciton fraction), condensation occurs in the lower branch; when the lower polariton is exciton-like (high exciton fraction), the upper polariton is the more photonic branch and condensation occurs there instead. 
This plot does not include data for very low temperature, where the condensate is in the lower branch; we cannot observe the UP branch at those low temperatures, because there is not sufficient thermal occupation to give PL from the upper state.

We emphasize that the condensates represented in Fig.~\ref{fig:phase_diagram} remain in the strong-coupling regime across these conditions. To rule out the possbility that the two lines in our measurements came from different regions, we performed spatially resolved measurements isolating the homogeneous center of the pump spot; this is the collection geometry used for the spectra shown in Fig. 1; additional measurements, together with the real-space images and the pinhole sizes, are given in the SI file. Even when the PL collection is restricted to only the very center of the condensate, the lower polariton branch is seen to retain its characteristic curvature and coexists with the upper polariton condensate, confirming that strong coupling persists at the condensation threshold.

\section{DISCUSSION}
Although the UP state is disfavored for condensation because it is not the
ground state, if the conversion between the populations of the UP and LP states
is slow, and the UP state has a lighter effective mass, the quantum degeneracy of
the UP state can be greater than that of the LP, allowing condensation to appear there first. This is an example of a transition in an overall nonequilibrium system that nevertheless has two populations each in quasiequilibrium. 

A simple model predicts whether a condensate will appear first in the LP or UP
state. We assume that both branches are populated from a common hot
electron--hole reservoir with density $n_{\rm res}$ and that particles can
interconvert between the two branches. The corresponding densities then obey
\begin{align}
\frac{d n_{\rm res}}{dt}
&= P-\frac{n_{\rm res}}{\tau_{\rm res}},
\label{eq:res}\\[4pt]
\frac{d n_{\rm UP}}{dt}
&= \frac{n_{\rm res}}{\tau^{R}_{\rm UP}}
-\frac{n_{\rm UP}}{\tau_{\rm UP}}
-\frac{n_{\rm UP}}{\tau_{\rm conv}}+\frac{n_{\rm LP}}{\tau_{\rm conv}}
\frac{\mathrm{DOS}_{\rm UP}}{\mathrm{DOS}_{\rm LP}}
e^{-\Delta E_0/k_B T},
\label{eq:up}\\[4pt]
\frac{d n_{\rm LP}}{dt}
&= \frac{n_{\rm res}}{\tau^{R}_{\rm LP}}
-\frac{n_{\rm LP}}{\tau_{\rm LP}}
+\frac{n_{\rm UP}}{\tau_{\rm conv}}-\frac{n_{\rm LP}}{\tau_{\rm conv}}
\frac{\mathrm{DOS}_{\rm UP}}{\mathrm{DOS}_{\rm LP}}
e^{-\Delta E_0/k_B T}.
\label{eq:lp}
\end{align}
Here, $P$ is the pump rate, $\tau^{R}_{\rm UP}$ and $\tau^{R}_{\rm LP}$ are the
times for the reservoir to fill each branch, $\tau_{\rm UP}$ and $\tau_{\rm LP}$
are the branch lifetimes, $\tau_{\rm conv}$ is the interconversion time,
$\mathrm{DOS}_{i}$ is the density of states of branch $i$, and
$\Delta E_0\equiv E_{\rm UP}(0)-E_{\rm LP}(0)$. Particle number conservation
requires that $1/\tau_{\rm res}=1/\tau^{R}_{\rm UP}+1/\tau^{R}_{\rm LP}$.

To calculate $\Delta E_0$ and the densities of states of the two species, we
assume a simple two-level system for the polaritons \cite{deng2010exciton}, with
a coupling Hamiltonian given by
\begin{equation}
H(k=0)
=
\left(
\begin{array}{cc}
E_{\rm phot}(0) & \Omega/2 \\
\Omega/2 & E_{\rm exc}(0)
\end{array}
\right),
\label{matrix}
\end{equation}
where $E_{\rm phot}(0)$ and $E_{\rm exc}(0)$ are the bare cavity photon and bare
exciton energies at $k=0$, respectively, and $\Omega$ is the Rabi coupling
energy. Diagonalization of this matrix gives
\begin{equation}
\label{eq:deltaE}
\Delta E_0
=
\sqrt{\Delta_0^{\,2}+\Omega^2}
=
\frac{\Omega}{2\sqrt{f_{\rm ex}(1-f_{\rm ex})}},
\end{equation}
where $f_{\rm ex}\equiv|X^{\rm LP}|^2$ is the exciton fraction of the LP, defined
from the eigenvectors of the Hamiltonian through the Hopfield coefficients
\cite{hopfield1958theory}, and the detuning is
\begin{equation}
\Delta_0
=
\Omega\,
\frac{2f_{\rm ex}-1}
{2\sqrt{f_{\rm ex}\left(1-f_{\rm ex}\right)}}.
\label{eq:detuning}
\end{equation}

The effective masses are deduced from (\ref{matrix}) when an additional term
$\hbar^2k_{\|}^2/2m_{\rm cav}$ is added to $E_{\rm phot}(0)$, where $m_{\rm cav}$
is the effective mass of the bare cavity photon mode. In the limit
$m_{\rm ex}\gg m_{\rm cav}$,
\begin{equation}
\frac{1}{m_{\rm LP}}\approx\frac{|C^{\rm LP}|^2}{m_{\rm cav}},
\qquad
\frac{1}{m_{\rm UP}}\approx\frac{|X^{\rm LP}|^2}{m_{\rm cav}},
\label{eq:masses}
\end{equation}
while the branch lifetimes are set by the photon fraction,
\begin{equation}
\tau_i=\frac{\tau_c}{|C_i|^2},
\label{eq:lifetimes}
\end{equation}
with $\tau_c$ the cavity lifetime. In two dimensions, the density of states is
independent of energy and proportional to the effective mass,
$\mathrm{DOS}_i=m_i/(2\pi\hbar^2)$, so that, using
$|C^{\rm UP}|^2=|X^{\rm LP}|^2$,
\begin{equation}
\frac{\mathrm{DOS}_{\rm UP}}{\mathrm{DOS}_{\rm LP}}
=
\frac{m_{\rm UP}}{m_{\rm LP}}
=
\frac{|C^{\rm LP}|^2}{|C^{\rm UP}|^2}
=
\frac{1-f_{\rm ex}}{f_{\rm ex}}.
\label{eq:dosratio}
\end{equation}
Equations (\ref{eq:masses}) and (\ref{eq:lifetimes}) also give
$\mathrm{DOS}_{i}k_BT/\tau_i=m_{\rm cav}k_BT/2\pi\hbar^2\tau_c$ for both
branches: the more photonic branch decays faster, but has proportionally fewer
states, and the two effects cancel.

The density of each branch follows from its chemical potential, given by
$\mathrm{DOS}_{i}\,k_B T\,\exp{[(\mu_i-E_i)/k_B T]},$ where $\mu_i$ is the
chemical potential of branch $i$. Because the interconversion is slow, the two
branches are not in mutual equilibrium and $\mu_{\rm UP}\neq\mu_{\rm LP}$; each
population has its own chemical potential. We absorb the exponential into a
single dimensionless variable, $N_i \equiv \exp{[(\mu_i-E_i)/k_B T]}$, which is
the occupation number of the states at the bottom of branch $i$, so that
$n_i=\mathrm{DOS}_{i}k_B T\,N_i$. Substituting into Eqs.~(\ref{eq:up}) and
(\ref{eq:lp}), the steady-state equations are
\begin{align}
0
&=
\frac{n_{\rm res}}{\tau^{R}_{\rm UP}}
-
\frac{m_{\rm cav}k_B T}{2\pi\hbar^2\tau_c}N_{\rm UP}
\nonumber\\
&\qquad
-
\frac{\mathrm{DOS}_{\rm UP}k_B T}{\tau_{\rm conv}}
\left(
N_{\rm UP}
-
N_{\rm LP}e^{-\Delta E_0/k_B T}
\right),
\label{eq:ss1}\\[4pt]
0
&=
\frac{n_{\rm res}}{\tau^{R}_{\rm LP}}
-
\frac{m_{\rm cav}k_B T}{2\pi\hbar^2\tau_c}N_{\rm LP}
\nonumber\\
&\qquad
+
\frac{\mathrm{DOS}_{\rm UP}k_B T}{\tau_{\rm conv}}
\left(
N_{\rm UP}
-
N_{\rm LP}e^{-\Delta E_0/k_B T}
\right).
\label{eq:ss2}
\end{align}
This is a linear system whose solution is
\begin{equation}
\frac{N_{\rm UP}}{N_{\rm LP}}
=
\frac{e^{-\Delta E_0/k_B T}+p r}
{1+(1-p)r},
\label{eq:master}
\end{equation}
which is independent of the reservoir density $n_{\rm res}$, with the two
dimensionless parameters
\begin{equation}
r
\equiv
\frac{\tau_{\rm conv}}{\tau_{\rm UP}},
\qquad
p
\equiv
\frac{\tau^{R}_{\rm LP}}
{\tau^{R}_{\rm UP}+\tau^{R}_{\rm LP}},
\label{eq:rp}
\end{equation}
where $r$ is the ratio of the loss and conversion coefficients in
Eqs.~(\ref{eq:ss1}) and (\ref{eq:ss2}), and $p$ is the fraction of the reservoir
converted to the upper branch.

The implications of Eq.~(\ref{eq:master}) can be understood by first considering
the two limiting cases. For fast interconversion, $r\to0$, it reduces to the
equilibrium case
\begin{equation}
\frac{N_{\rm UP}}{N_{\rm LP}}
=
e^{-\Delta E_0/k_B T}
<1.
\end{equation}
The lower branch always wins, the pumping drops out entirely, and upper-branch
condensation is impossible. In the opposite limit, where the two species are
completely decoupled and there is no exchange between them, $r\to\infty$, the
energy offset drops out and
\begin{equation}
\frac{N_{\rm UP}}{N_{\rm LP}}
\longrightarrow
\frac{\tau^{R}_{\rm LP}}{\tau^{R}_{\rm UP}}.
\end{equation}
Every Hopfield coefficient has canceled, and the outcome is set purely by which
branch the reservoir fills faster. The present experiment lies between these two
limits, which is why a sample with $\tau_c$ comparable to $\tau_{\rm conv}$ was
used.

The crossover between UP and LP condensation occurs when the degeneracy ratio in
Eq.~(\ref{eq:master}) equals unity. Setting $N_{\rm UP}/N_{\rm LP}=1$ gives
\begin{equation}
r(2p-1)
=
1-e^{-\Delta E_0/k_B T}.
\label{eq:criterion}
\end{equation}
Condensation in the upper branch therefore requires both $p>1/2$ and
\begin{equation}
r
>
\frac{1-e^{-\Delta E_0/k_B T}}{2p-1},
\end{equation}
which is consistent with the quantum Boltzmann simulations reported in
Ref.~\cite{chen2023bose}, where UP condensation is achieved only when the UP
branch is pumped more strongly than the LP branch. The reservoir must fill the
upper branch faster than the lower one, and the interconversion must be slow
enough not to drain it again. Combining the crossover condition
(\ref{eq:criterion}) with the splitting (\ref{eq:deltaE}) and using
$\tau_{\rm UP}=\tau_c/f_{\rm ex}$ from (\ref{eq:lifetimes}), the phase boundary
follows in closed form:
\begin{equation}
T(f_{\rm ex})
=
-\frac{\Omega}
{2k_B\sqrt{f_{\rm ex}(1-f_{\rm ex})}\,
\ln\!\left(1-qf_{\rm ex}\right)},
\label{eq:boundary}
\end{equation}
where
\begin{equation}
q
\equiv
\frac{\tau_{\rm conv}}{\tau_c}(2p-1).
\end{equation}
This is the solid line in Fig.~\ref{fig:phase_diagram}.

Equation~(\ref{eq:boundary}) incorporates the tendency for condensation to occur in the lower-energy state while also allowing nonequilibrium filling to favor the upper branch. Thermal interconversion always favors the LP through the Boltzmann factor $e^{-\Delta E_0/k_B T}$, which remains smaller than unity. In contrast, preferential filling of the UP can allow it to reach quantum degeneracy first if the reservoir fills the UP faster than the LP and the interconversion is sufficiently slow. At low temperatures, however, the Boltzmann factor becomes exponentially small, and the condensate will always appear in the LP branch unless the nonequilibrium filling imbalance is large enough to compensate for this energy difference. The competition between relaxation toward the lower LP energy and preferential filling of the UP branch defines the phase boundary in Fig.~\ref{fig:phase_diagram}.

The data (e.g., Figure~\ref{fig:condensation}) shows that the Rabi splitting
$\Omega$ decreases as the density increases, as expected due to Pauli phase-space
filling (see, e.g.,
\cite{rhee1995nonlinear,houdre1995saturation,schmitt1985theory}).

Since higher temperature requires high density for condensation, approximately
given by $n_c = m k_B T / \hbar^2$ in two dimensions, we assume the Rabi coupling
that enters into (\ref{eq:deltaE}) has form $\Omega = \Omega_0(1 - \alpha n_c)$,
where $n_c$ is computed using an average effective mass
$m = (1/m_{\mathrm{LP}} + 1/m_{\mathrm{UP}})^{-1} = m_{\mathrm{cav}}$.

The model has two fit parameters, which are $q$ and $\alpha$. The zero-density
splitting and the effective mass of the bare cavity photon are directly measured
and are given by $\Omega_0=12\;\mathrm{meV}$ and
$m_{\rm cav}=4\times10^{-5}m_e$. The solid line in Fig.~\ref{fig:phase_diagram}
corresponds to $q=1.2$ and $\alpha=0.037\;\mathrm{\mu m^2}$. With a cavity
lifetime $\tau_c\simeq20\;\mathrm{ps}$, $q=1.2$ implies
$\tau_{\rm conv}(2p-1)\simeq24\;\mathrm{ps}$, of the same order as the cavity
lifetime. As seen in the figure, the predicted phase boundary gives quite good
agreement with the experimental results. Note that although Fig.~\ref{fig:phase_diagram} does not show data from other experiments, all prior experiments using microcavity samples with
this design show only LP condensation when $T<50$~K, consistent with
Eq.~(\ref{eq:criterion}) since $\exp{(-\Delta E_0/k_B T)}\to0$ in that limit.

\section{CONCLUSIONS}

We have demonstrated temperature-dependent switching of polariton condensation between two different species of polaritons in a GaAs/AlGaAs microcavity. Remarkably, a simple model of the relative degeneracy of the two states predicts the phase diagram of this switching, with the driven-dissipative dynamics entering only through the single dimensionless parameter $q$, which combines the interconversion time, the cavity lifetime, and the fraction of the reservoir that feeds each branch. Essentially, when the two mixed species have slow interconversion and roughly equal population, whichever has the lighter mass will condense, because it will have greater quantum degeneracy.

Near 80 K, the system exhibits an instability in condensate formation, with stochastic switching between branches under the same pumping power. This points us to explore the critical dynamics near the phase transition leading to mode switching, similar to that seen in prior work with much lower energy difference between the stable modes \cite{alnatah2024critical,rozas2026temporal}.

The behavior we see indicates that the switching between the two condensate states is hypersensitive to the parameters of the experiment when the system is at the phase boundary. Therefore, this phenomenon could potentially be adapted for nonlinear switching applications, for example, using short, weak laser pulses to push the system over the phase boundary between UP and LP condensation.

\section{Acknowledgments}
This project has been supported by the National Science Foundation grant  DMR-2306977. We thank M. Szymanska and P. Comaron for helpful conversations.

\clearpage
\bibliography{references.bib}

\clearpage
\date{\today}
\maketitle
\beginsupplement
\section{Sample design}

\begin{figure}[H]
\includegraphics[width=0.9\columnwidth]{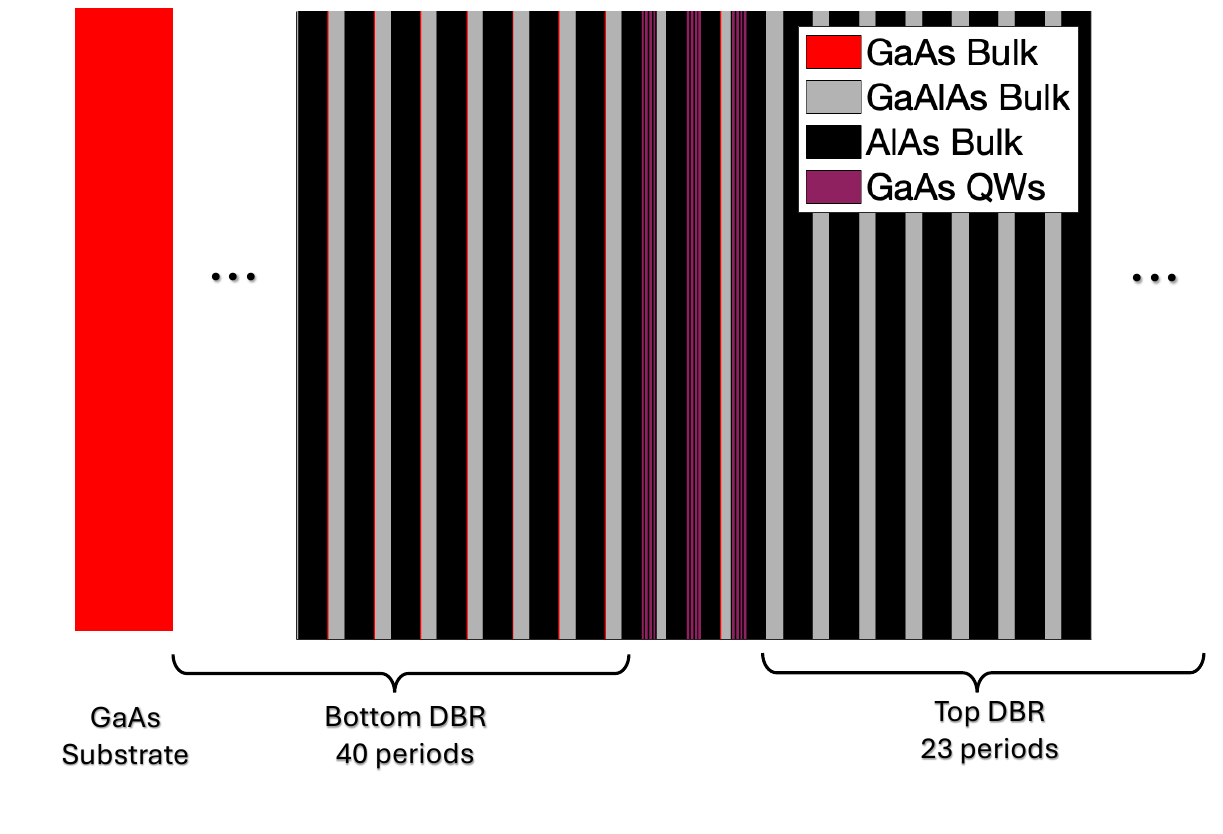}
\centering
\caption{\textbf{Schematic illustration of the sample structure.} The GaAs quantum wells are embedded in a GaAs/AlGaAs microcavity formed by DBRs.}
\label{fig:sample}
\end{figure}

The sample used in this work, as shown in Fig.~\ref{fig:sample}, contained a total of 12 GaAs quantum wells (QWs) separated by AlAs barriers and embedded within a distributed Bragg reflector (DBR) microcavity with 23 pairs in the top DBR. All quantum wells had the same thickness and were arranged in groups of four, with each group positioned at an antinode of the cavity field. Furthermore, the DBRs consisted of alternating layers of AlAs and \ce{Al_{0.2}Ga_{0.8}As}.

\section{Characteristics of PL lines}

\begin{figure}[H]
\includegraphics[width=0.75\columnwidth]{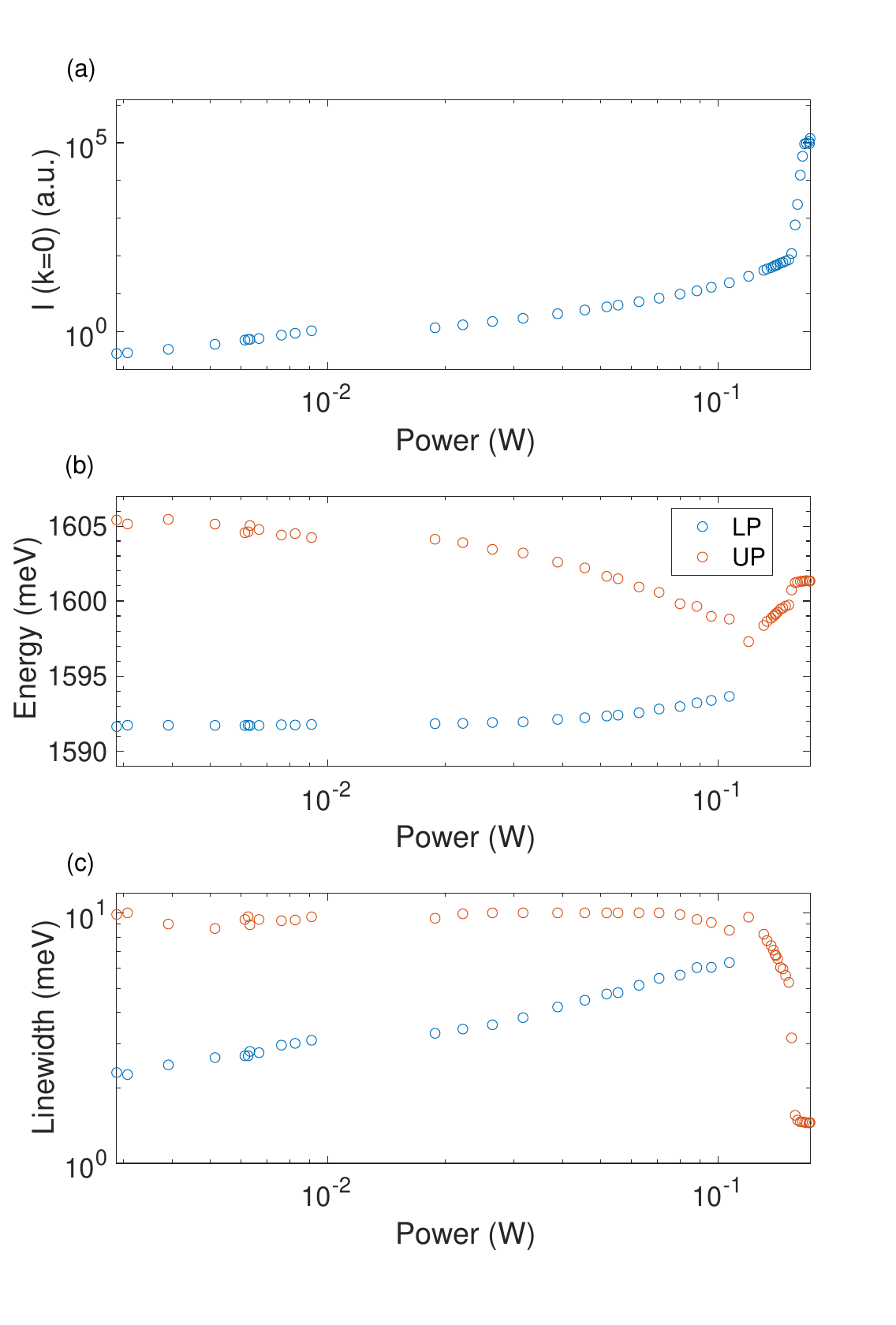}
\centering
\caption{\textbf{Characteristics of the PL lines at 100 K and an exciton fraction of 0.61 corresponding to Fig. 1(a-b) in the main text.}\textbf{(a)} The intensity at $k=0$ of the polaritons as a function of pump power. \textbf{(b)} The energies of the polariton lines at $k=0$ as a function of the pump power. \textbf{(c)}  Full width at half maximum at $k=0$. }
\label{fig:energy_linewidth_100K}
\end{figure}

\begin{figure}[H]
\includegraphics[width=0.75\columnwidth]{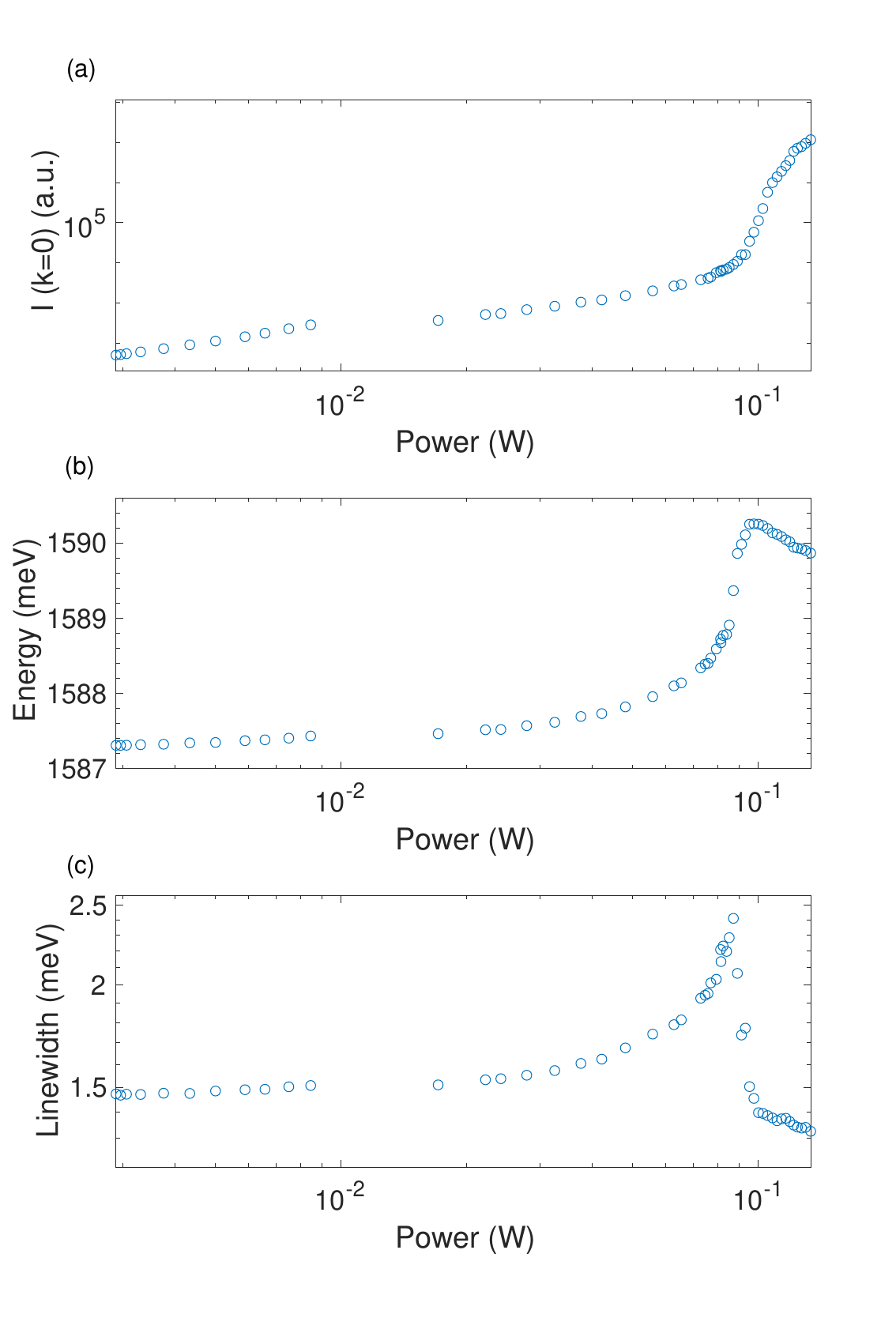}
\centering
\caption{\textbf{Characteristics of the PL lines at 80 K and an exciton fraction of 0.27 corresponding to Fig. 1(c-d) in the main text.} \textbf{(a)} The intensity at $k=0$ of the lower polariton as a function of pump power. \textbf{(b)} The energies of the lower polariton lines at $k=0$ as a function of the pump power. \textbf{(c)}  Full width at half maximum at $k=0$.}
\label{fig:energy_linewidth_70K}
\end{figure}

Figures~\ref{fig:energy_linewidth_100K} and \ref{fig:energy_linewidth_70K} show the power-dependent characteristics of the polariton emission at 100 K and 80 K, respectively. A clear threshold-like increase in the k=0 emission intensity is observed in Fig.~\ref{fig:energy_linewidth_100K}(a). The UP branch exhibits a pronounced linewidth narrowing together with a characteristic change in its energy shift (see Figs. S2(b,c)), which are signatures commonly associated with polariton condensation. These data correspond to the measurements shown in Fig. 1(a,b) of the main text, obtained at 100 K, where the LP exciton fraction is approximately 0.65.
Figure \ref{fig:energy_linewidth_70K} shows the power dependence of the LP photoluminescence at k=0. These data correspond to the same dataset shown in Fig. 1(c,d) of the main text, obtained at 80 K, where the LP exciton fraction is approximately 0.27. Because the upper-polariton (UP) mode at k=0 is predominantly excitonic at this detuning and therefore exhibits much weaker photoluminescence than the more photonic LP mode, only the LP emission is analyzed here. The LP emission exhibits a gradual blueshift together with linewidth reduction at elevated pump powers, indicating the onset of LP condensation. Thus, at 80 K the nonlinear behavior is associated with the LP branch rather than the UP branch. This behavior contrasts with that observed at 100 K, where condensation signatures predominantly appear in the UP branch, further supporting the temperature-dependent switching between LP and UP condensation.

\section{Spatial filtering}
To address the concern that the emission at high density may originate from a transition to weak coupling (e.g. cavity phonon lasing) at the center of the pump spot—while strong coupling persists only in the outer regions—we performed spatially resolved measurements using a pinhole to isolate emission only from the pump center in a region that is spatially homogeneous.
\par
As shown in Fig.~\ref{fig:EvsK_condensation}(a–b), the PL was selected in real space using a pinhole of diameter $40\,\mu\mathrm{m}$. The pinhole is not perfectly centered, as it was aligned with the brightest region of the condensate at the highest excitation power. Nevertheless, the intensity within the selected region remains sufficiently homogeneous.

With this spatial filtering, we clearly observe that the lower polariton branch remains visible even when upper polariton condensation occurs (see Fig.~\ref{fig:EvsK_condensation}(d)). The data are plotted on a logarithmic scale to emphasize the lower polariton emission, since the upper polariton condensate is significantly brighter.
\begin{figure}[H]
\includegraphics[width=1\columnwidth]{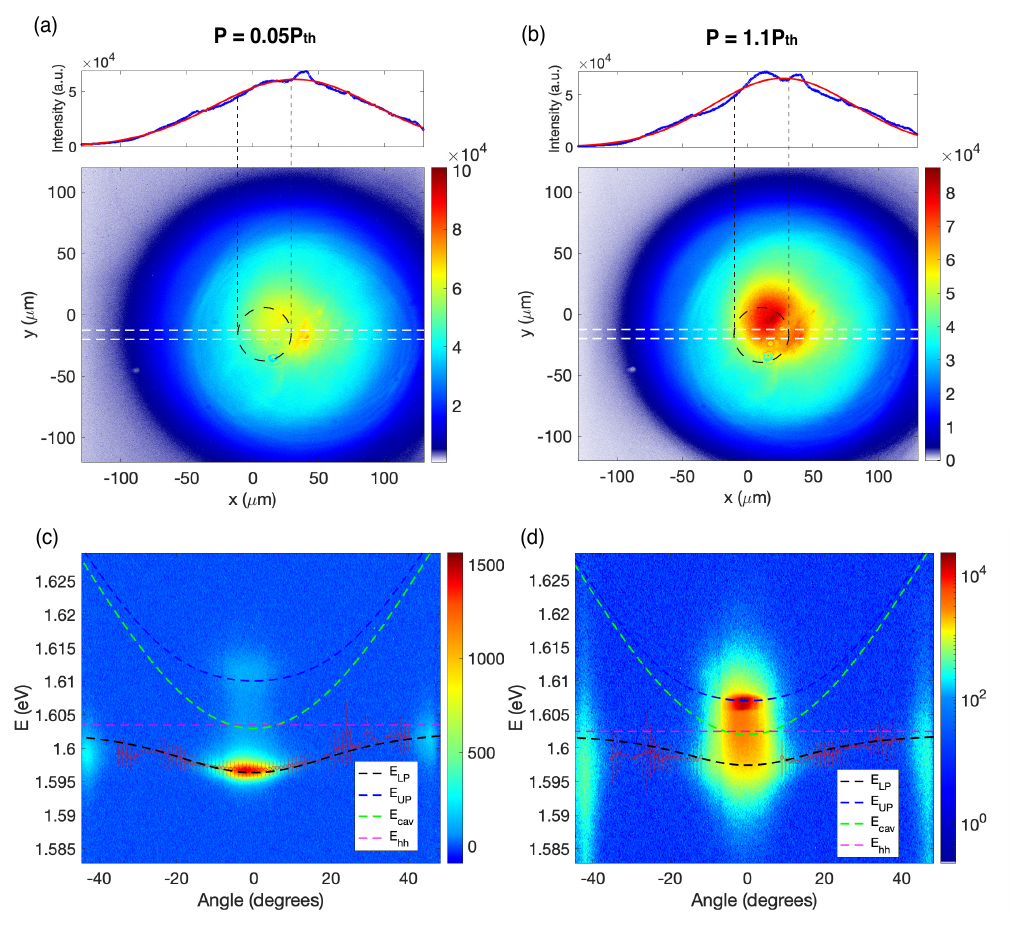}
\centering
\caption{\textbf{Spatially and angle-resolved photoluminescence at 80 K for an exciton fraction of the lower polariton is 0.72.} {\bf (a)} Real-space emission of polaritons at very low pumping power and {\bf (b)} at high pumping power at 80 K. The black dashed circle indicates the position of the pinhole, with a diameter of approximately $40\,\mu\mathrm{m}$. {\bf(c,d)} Angle-resolved photoluminescence at the corresponding low and high pumping powers, respectively, after spatial filtering through the pinhole. The red circles represent the extracted dispersion, obtained by fitting each angle-resolved linecut with a Lorentzian. The dashed lines show fits to a two-level exciton-polariton model: the black line denotes the lower polariton branch, the blue line the upper polariton branch, the green line the cavity mode, and the magenta line the heavy-hole exciton.}
\label{fig:EvsK_condensation}
\end{figure}
Importantly, the lower polariton branch retains a finite curvature rather than appearing as a straight line, confirming that the system remains in the strong coupling regime. The markers shown in Fig.~\ref{fig:EvsK_condensation}(c,d) are obtained by taking energy spectra at different emission angles and fitting the corresponding intensity $I(E)$ with a Lorentzian function to extract the peak positions. These extracted points clearly demonstrate that the lower polariton branch persists and coexists with the upper polariton condensate.

\begin{figure}[H]
\centering
\includegraphics[width=1\columnwidth]{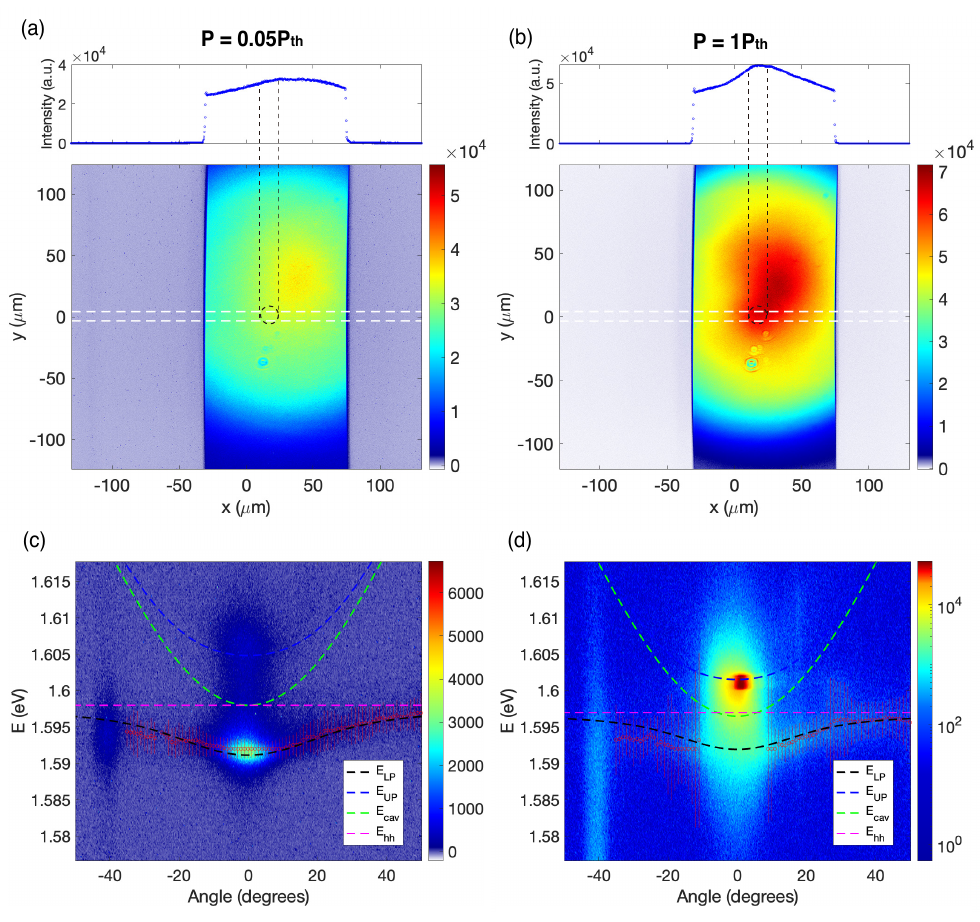}
\caption{\textbf{Spatially and angle-resolved photoluminescence at 100 K for an exciton fraction of the lower polariton is 0.65.} {\bf (a)} Real-space emission of polaritons at very low pumping power and {\bf (b)} at high pumping power at 100 K. The black dashed circle indicates the position of the pinhole, with a diameter of approximately $10\,\mu\mathrm{m}$. {\bf(c,d)} Angle-resolved photoluminescence at the corresponding low and high pumping powers, respectively, after spatial filtering through the pinhole. The red circles represent the extracted dispersion, obtained by fitting each angle-resolved linecut with a Lorentzian. The dashed lines show fits to a two-level exciton-polariton model: the black line denotes the lower polariton branch, the blue line the upper polariton branch, the green line the cavity mode, and the magenta line the heavy-hole exciton.}
\label{fig:EvsK_condensation_100K}
\end{figure}

To confirm that this behavior is not specific to a single temperature or detuning, we repeated the measurement at 100 K for a lower-polariton exciton fraction of 0.65 (Fig.~\ref{fig:EvsK_condensation_100K}). A smaller pinhole of $10\,\mu\mathrm{m}$ diameter, positioned at the location indicated by the black dashed circle in Fig.~\ref{fig:EvsK_condensation_100K}(a,b), was used to isolate the center of the pump spot. (The image appears truncated on both sides due to the spectrometer entrance slit.)

As at 80 K, the lower polariton branch remains clearly resolved both at low excitation power and at the condensation threshold, and it retains the finite curvature characteristic of the strong-coupling regime. The condensate again forms on the upper polariton branch (Fig.~\ref{fig:EvsK_condensation_100K}(d)). 

\begin{figure}[H]
\centering
\includegraphics[width=1\columnwidth]{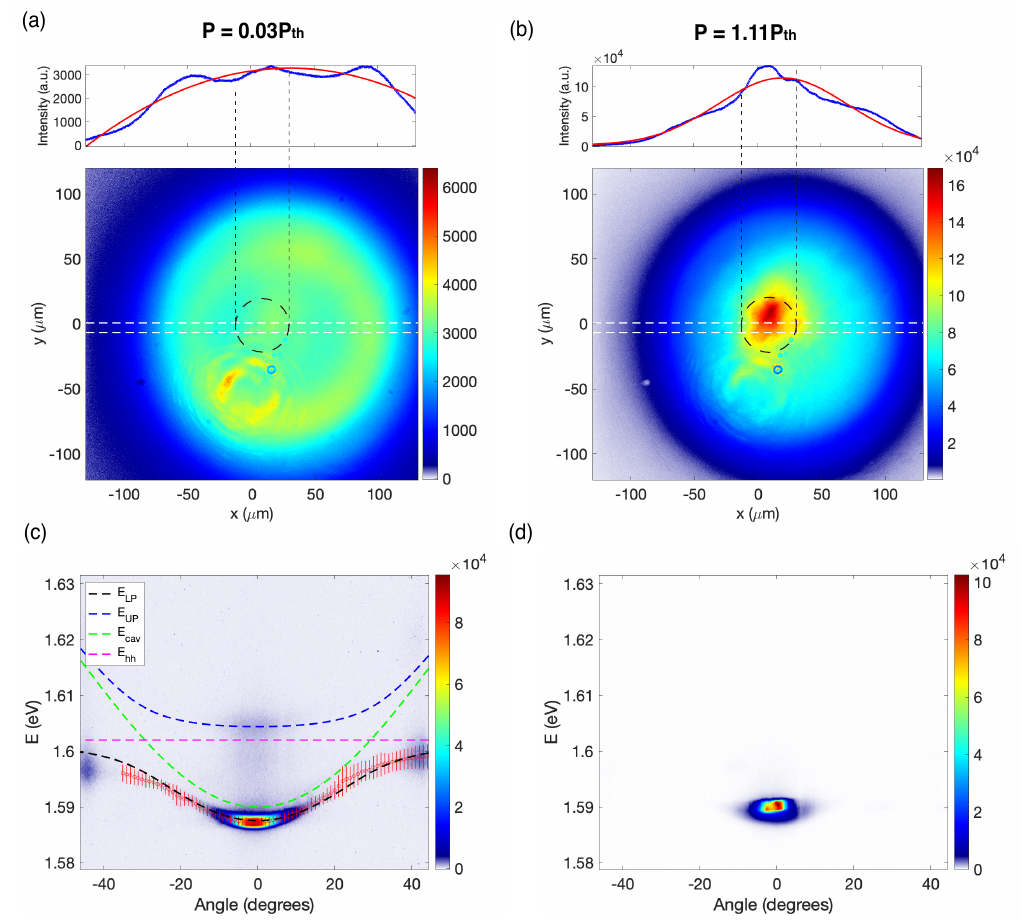}
\caption{\textbf{Spatially and angle-resolved photoluminescence at 80 K for an exciton fraction of the lower polariton is 0.27.} {\bf (a)} Real-space emission of polaritons at very low pumping power and {\bf (b)} at high pumping power at 80 K. The black dashed circle indicates the position of the pinhole, with a diameter of approximately $40\,\mu\mathrm{m}$. {\bf(c,d)} Angle-resolved photoluminescence at the corresponding low and high pumping powers, respectively, after spatial filtering through the pinhole. The red circles represent the extracted dispersion, obtained by fitting each angle-resolved linecut with a Lorentzian. The dashed lines show fits to a two-level exciton-polariton model: the black line denotes the lower polariton branch, the blue line the upper polariton branch, the green line the cavity mode, and the magenta line the heavy-hole exciton.}
\label{fig:S6}
\end{figure}

\end{document}